\documentclass{emulateapj}
\usepackage{apjfonts}

\newcommand{\etal}{et al.~}
\def\gsim{\lower 2pt \hbox{$\, \buildrel {\scriptstyle >}\over
{\scriptstyle \sim}\,$}}
\def\lsim{\lower 2pt \hbox{$\, \buildrel {\scriptstyle <}\over
{\scriptstyle \sim}\,$}}
\def\rosat{{\sl ROSAT}}

\def\chandra{{\sl Chandra~}}

\def\oviii{O~{\scriptsize VIII}}
\def\ovii{O~{\scriptsize VII}}
\def\ovi{O~{\scriptsize VI}}

\def\xs{GCSXE}

\shortauthors{Yao \& Wang}
\shorttitle{Galactic Central Diffuse X-Ray Enhancement}

\begin{document}

\slugcomment{\em Accepted for publication in the  Astrophysical Journal}

\title{The Galactic Central Diffuse X-Ray Enhancement: A Differential Absorption/Emission Analysis}
\author{Yangsen Yao\altaffilmark{1} and Q. Daniel. Wang\altaffilmark{2}}
\altaffiltext{1}{Massachusetts Institute of Technology (MIT) Kavli Institute 
 for Astrophysics and Space Research, 70 Vassar Street, Cambridge, MA 02139; 
 yaoys@space.mit.edu}
\altaffiltext{2}{Department of Astronomy, University of Massachusetts, 
  Amherst, MA 01003; wqd@astro.umass.edu}

\begin{abstract}

The soft X-ray background shows a general enhancement
toward the inner region of the Galaxy. But whether this enhancement is
a local feature (e.g., a superbubble within a distance of $\lsim 200$ pc) 
and/or a phenomenon related to energetic outflows from the Galactic center/bulge 
remains unclear. Here we report a comparative X-ray emission and 
absorption study of diffuse hot gas along the sight 
lines toward 3C~273 and Mrk~421, on and off the enhancement, but at similar
Galactic latitudes. The diffuse 3/4-keV emission intensity, as estimated from
the \rosat\ All Sky Survey, is about 3 times higher toward 3C~273 
than toward Mrk~421.  
Based on archival \chandra\ grating  observations of these two AGNs, 
we detect X-ray absorption lines (e.g., \ovii\ K$\alpha$, K$\beta$,
and \oviii\ K$\alpha$ transitions at $z \sim 0$) and find that the mean hot gas
thermal and kinematic properties along the two sight lines are significantly 
different. By subtracting the foreground and background contribution, 
as determined along the Mrk~421 sight line, 
we isolate the net X-ray absorption and emission produced by the hot gas 
associated with the enhancement in the direction of 3C~273. 
From a joint analysis of these differential data sets, we obtain the 
temperature, dispersion velocity, and hydrogen column density 
as $2.0(1.6, 2.3)\times10^6$ K, 216(104, 480)~km~s$^{-1}$, and
$2.2(1.4, 4.1)\times10^{19}~{\rm cm^{-2}}$, respectively (90\% confidence
intervals), assuming that the gas is approximately isothermal, solar
in metal abundances, and  equilibrium in collisional ionization. 
We also constrain the effective line-of-sight extent of the gas to be 
3.4(1.0, 10.1) kpc, strongly suggesting that the enhancement 
most likely represents a Galactic central phenomenon.

\end{abstract}

\keywords{Galaxy: structure --- X-rays: ISM --- X-rays: individual (3C~273, Mrk 421)}

\section{Introduction }
\label{sec:intro}
\begin{figure}
\centerline{
\plotone{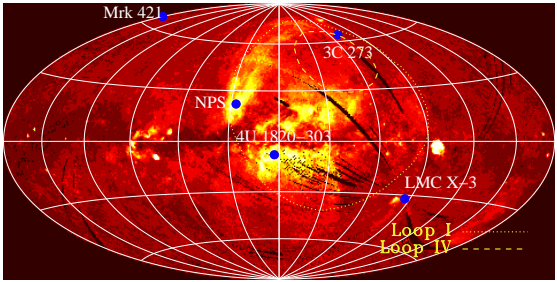}}
\caption{
  Directions of various relevant point sources (for X-ray absorption
 measurements) as well as the diffuse features: NPS and the radio loops I/IV 
(Haslam \etal 1971), shown on the RASS 3/4 keV band intensity 
  map \citep{sno97} in Aitoff projection.
  \label{fig:RASS} }
\end{figure}

The soft X-ray background (SXB) is greatly enhanced toward the inner part 
of the Galaxy ($l \lesssim 60^\circ$), particularly obvious in the 
\rosat\ All-Sky Survey (RASS) 
3/4-keV band map (Fig.~\ref{fig:RASS}; Snowden et al. 1997). 
This Galactic central soft X-ray 
enhancement (GCSXE) apparently arises from diffuse hot 
gas; but its three-dimensional (3-D) morphology and origin are largely uncertain.
The most prominent part of the \xs\  is along a ridge that is commonly 
considered to be associated with the North Polar Spur (NPS) --- 
probably part of the radio continuum feature Loop I. 
The NPS is believed to be a local feature, based primarily on
stellar polarization measurements (Bingham 1967); but this assertion is
not universally accepted. The NPS and the rest of Loop I 
have been interpreted as a nearby ($D \sim 170$ pc) supernova remnant 
(SNR; e.g., Iwan 1980), a stellar wind bubble from the Scorpio-Centaurus 
OB association, or a combination of the two (e.g., Egger \& Aschenbach 1995).
Alternatively, the \xs\  could represent a recent outflow (e.g., a bipolar 
wind) or a giant explosion from the Galactic nuclear region (Sofue 1984; 
Bland-Hawthorn \& Cohen 2003), or a more gentle but lasting Galactic bulge 
wind interacting with the Galactic gaseous halo (\S~\ref{sec:dis}).
In this case, the physical size of the enhancement must be
comparable to our distance to the Galactic center ($\sim 8$ kpc).
Clearly, these scenarios have vastly different implications for
the overall energetics of the \xs\ (i.e., $10^{51}$ vs. 
$\gtrsim 10^{54}$ ergs) and its impact on the global 
interstellar medium (ISM) structure of 
the Galaxy. 

Studies of the \xs\ have been based primarily on broadband X-ray 
observations such as the RASS (Snowden et al. 1997) and more recently on
X-ray CCD observations from {\sl XMM-Newton} and {\sl Suzaku}. 
By observing the X-ray intensity variation on and off cool gas clouds 
at known distances, one can separate the foreground and background 
contributions (Burrows \& Mendenhall 1991; Wang \& Yu 1995; Park \etal 1997;
Almy \etal 2000; Kuntz \& Snowden 2000; Galeazzi \etal 
2007; Smith \etal 2006; Henley \etal 2007). These experiments are particularly
useful in isolating the local X-ray emission and hence determining the 
properties of hot gas in the solar neighborhood. However, to infer
the properties of hot gas in distant regions is a much bigger challenge.
Particularly uncertain is the correction for the absorption by
intervening cool gas; the distribution of which relative to the X-ray-emitting
hot gas can be a complex (e.g, Kuntz \& Snowden 2000; 
Breitschwerdt \& de Avillez 2006). 
Nevertheless, it has been shown that
a large amounts of hot gas is present in regions toward the Galactic inner
region and beyond $d \sim 2 $ kpc (Almy et al. 2000). Along such a line of 
sight, one in general expects X-ray emission 
contributions from multiple components: 
the Local Hot Bubble (LHB; e.g., Snowden \etal 1998), 
the extended hot Galactic disk/corona 
(e.g., Yao \& Wang 2005, 2007), the 
unresolved Galactic and extragalactic point sources 
(e.g., Kuntz \& Snowden 2001; Hickox \& Markevitch 2006),
and the \xs. Of course, there could be additional discrete features
such as the NPS, which is most likely a separate identity, independent of
the \xs\ (see \S~4). Direct spectral decomposition of these components
along a single sight line is very difficult, if not impossible.

In this work, we present the first X-ray absorption line 
spectroscopy of the \xs, based on grating observations from
the {\sl Chandra X-Ray Observatory}. We compare the observations of 3C~273 and 
Mrk 421, on and off the \xs\ (Fig.~\ref{fig:RASS})
to determine its \ovii\ and \oviii\ line absorptions. This
differential spectroscopic data set enables us to characterize the thermal and 
kinematic properties of the hot gas associated with the \xs\ with
minimal uncertainties. The spectroscopic data, together with the 
emission intensity difference measured with the RASS, further allow us to
estimate the effective path-length and density of the 
hot gas.

Throughout the paper, we quote the errors at the 90\% confidence level, adopt the solar metal abundance from \citet{and89},
and assume that the hot gas is in the collisional ionization
equilibrium (CIE) state and is approximately isothermal. 

\section{Observations and Data Reduction}

As shown in Figure~\ref{fig:RASS}, the sight line of 3C~273 passes 
through the \xs, and is at the edges of radio loops I/IV. 
The 3/4-keV (RASS R4+R5 band) background is known to have quite 
a uniform intensity distribution, except for those abnormal regions 
such as the Galactic Bulge, NPS, the Cygnus Loop, etc. \citep{sno97}.
Much of the background emission in this band is proved to be thermal,
with distinct line emission in high spectral resolution observations
\citep{mcc02}. In contrast, the 1/4-keV (R1+R2 band) background intensity 
traces mostly hot gas in the LHB and residual emission 
from more distant regions after significant photo-electric absorption by 
foreground cool ISM. The background intensity in the 1/4-keV band
appears patchy and shows little correlation with those in the
higher energy bands. The background intensity in the 1.5-keV (R6+R7) band 
arises primarily from extragalactic AGNs. Therefore we only use the data in
3/4-keV band that are primarily due to \ovii\ and \oviii\ line emission
in the hot ISM \citep{mcc02}.

The diffuse 3/4-keV background intensity
in the vicinity of the 3C~273 sight line is about 3 times of those at 
a typical high ($|b|\gsim20^\circ$) Galactic latitudes but away from 
the inner part of the Galaxy (e.g., Mrk~421 direction)
and is about 1/5 of that at the brightest part
of the NPS (Table~\ref{tab:ROSAT}; Snowden \etal 1997).
We use Mrk~421 direction as a reference sight line, which has a 
Galactic latitude similar to the 3C~273 direction.

\begin{deluxetable}{lcccc}[b]
\tablewidth{0pt}
\tablecaption{Properties of the interested fields\label{tab:ROSAT}}
\tablehead{
          &   & $N_{\rm HI}$ \\
objects          & ($l, b$)   &(${\rm cm^{-2}}$)& SXB Int. & Diffuse Int. }
\startdata
  3C~273 &($289\fdg95, ~64\fdg36$)&$1.7\times10^{20}$ &$1.92\times10^{-4}$ & $1.29\times10^{-4}$ \\
  Mrk~421&($179\fdg83, ~65\fdg03$)&$1.4\times10^{20}$ &$1.01\times10^{-4}$ & $0.39\times10^{-4}$ \\
  NPS    &($~26\fdg84, ~21\fdg96$)&$5.6\times10^{20}$ &$7.58\times10^{-4}$ & $~~6.95\times10^{-4}$ 
\enddata
\tablecomments{
The total HI absorption column density are 
   adopted from \citet{sem01} for 3C~273 and from Dickey \& Lockman (1990) 
   for Mrk~421 and NPS directions. 
   The SXB intensities (in units of ${\rm counts~s^{-1}~arcmin^{-2}}$)
   toward 3C~273 and Mrk~421 are obtained from an annulus of $1\fdg0$ 
   to $1\fdg5$ radii around the source on the 3/4-keV band RASS
   map. The diffuse intensities are obtained by subtracting the 
   contributions from the unresolved extragalactic and Galactic point sources
   (\S~\ref{sec:intro}). The intensities of the NPS (its brightest part) 
   are obtained from a circular region of $1^\circ$  radius.
}
\end{deluxetable}

\begin{figure}
\plotone{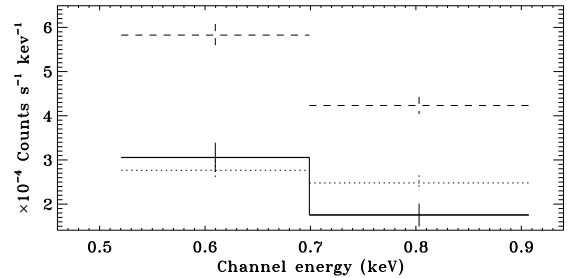}
\caption{Spectrum of the emission associated 
	with the \xs\  and in the direction of 3C~273 ({\sl solid crosses}), 
	the best-fit model 
	convolved with the RASS response ({\sl solid histogram}), 
	and the SXB spectra toward 3C~273 ({\sl dashed crosses}),
	and Mrk~421 ({\sl dotted crosses}). See text for details.
   \label{fig:ROSAT} }
\end{figure}

We use the RASS intensities in the R4 and R5 sub-bands to gain crude emission 
spectral characteristics of the 3C~273 and Mrk~421 sight lines. 
To account for the possible calibration uncertainties in the intensities
of the individual energy bands, we add 10\% 
\footnote{
The calibration uncertainty in intensities of the individual bands
is not available in the literature. The 10\% is our attempted value.}
systematic uncertainty to the emission spectrum in our data analysis.
To avoid the potential confusion from the 
emission of these bright sources due to the broad wing of the 
point spread function (PSF) of the RASS 
\footnote{For a PSPC pointing observation with an off-axis of $0'(30')$, the 
PSF density at 0.5 keV drops $\sim10^8(10^6)$ times at $\sim1^\circ$ away from 
the point source. For Mrk~421 at its historical high 
flux ($\sim50$ times higher than the normal flux; e.g., the {\sl Chandra} 
observation with ObsID 4148), the stray light from the point source 
contributes $\leq 0.2\%(20\%)$ to the SXB intensity of an annulus of 
$1\fdg0$ to $1\fdg5$ radii for R4+R5 band. The contribution from 3C~273 
to the SXB intensity is negligible.}
and to smooth out any potential local 
emission gradient, we extract the background intensities 
(Fig.~\ref{fig:ROSAT}) from an 
annulus of $1\fdg0$ to $1\fdg5$ radii centered at each of the sources, 
using the X-Ray Background Tool in HEASAC. 
\footnote{http://heasarc.gsfc.nasa.gov/cgi-bin/Tools/xraybg/xraybg.pl}

The absorption spectra are extracted from an extensive set of 
{\sl Chandra} grating observations, with an accumulated exposure time 
of $\sim480$ ks each for 3C~273 and Mrk~421. We have
analyzed most of the observations of Mrk~421 in \citet{yao07} and have 
detected significant \ovii\ K$\alpha$, K$\beta$, and \ovii\ K$\alpha$ 
absorption lines in a co-added spectrum. 
For this study of the 3C~273 sight line, we use 13 observations taken 
with the Advanced CCD Imaging Spectrometer (ACIS). Among these
observations (ObsID: 1198, 2464, 
2471, 3574, 4431, 5170, 459, 2463, 3456, 3457, 3573, 4430, 
and 5169), the last seven used the High Energy Transmission Grating 
(HETG), for which only the data from the medium-energy-grating arms are 
included. We do not use two observations taken with the High Resolution 
Camera for a total exposure of $\sim70$ ks; their inclusion would
increase the counting statistics only by $\sim15\%$ at $\sim$ 20 \AA, but 
would also introduce the complex overlapping from high ($>1$) order data
(Yao \& Wang 2006, 2007). We process the ACIS data, following the same 
procedure as described in Yao \& Wang (2006, 2007).

We focus on the oxygen absorption lines to avoid the uncertainty in 
dealing with relative metal abundances.
We fit the spectral continuum in the 18--22 \AA\ range
with a power law and fix the foreground cool gas absorption to be 
$N_H^c=1.7\times10^{20}~{\rm cm^{-2}}$
\citep{sem01}. This model fit gives $\chi^2$/dof = 524/397; the 
strong \ovii\ K$\alpha$, K$\beta$, and \oviii\ K$\alpha$
absorption lines are clearly visible in the residual.  
The inclusions of these absorption lines as described in
\S~3 significantly
improves the fit ($\chi^2/$dof $\simeq 402/392$). The resultant power-law 
index and the normalization are $2.1(2.0, 2.3)$ and
$2.3(2.2, 2.5)\times10^{-2}~{\rm photons~keV^{-1}~cm^{-2}~s^{-1}}$ 
at 1 keV. 

The \ovii\ K$\alpha$ absorption line  toward
3C~273 has already been detected by \citet{fang03}, based on a subset of 
the observations used in this work. They have further discussed possible 
origins of the absorbing hot gas as a single component, either the 
intergalactic medium, the Galactic halo, or the radio loops I/IV. 
The detection of the additional absorption lines and the
comparative analysis between the sight lines, as reported 
here, now make it possible to isolate and characterize the contribution
from the \xs.

\section{Analysis and Results}
\label{sec:results}
We first characterize the absorbing gas by using our absline 
model\footnote{See \citet{yao05} and \citet{wang05} for a detailed 
description of the model.} to jointly fit the \ovii\ and \oviii\ absorption 
lines in the spectra of 3C~273 and Mrk~421. We measure the dispersion 
velocity ($v_b$), temperature ($T$), and the effective hydrogen column 
density ($N_{\rm H}$) of the
gas along each sight line. Table~\ref{tab:abs} summarize the results,
including the inferred \ovi, \ovii, and \oviii\ column densities. 
This single-temperature modeling represents only a simplified 
characterization of the gas. For example, a joint analysis
of the emission and absorption data for Mrk 421 shows that the 
hot gas along the sight line is not isothermal
\citep{yao07}. We expect a more complicated temperature structure
along the 3C~273 sight line. Nevertheless, the simple characterization, 
insensitive to the 
specific physical model adopted, should be sufficient for the modeling of 
averaged hot gas properties along the two sight lines,  
and thus facilitates a differential analysis of the \xs. 

\begin{deluxetable*}{lcccccc}
\tabletypesize{\small}
\tablewidth{0pt}
\tablecaption{
  Spectral fit Results\label{tab:abs} }
\tablehead{
&$v_b$         & T   & N$_{\rm H}$ & N$_{\rm OVII}$ &N$_{\rm OVIII}$ & N$_{\rm OVI}$\\
direction & (km s$^{-1}$) & ($10^6$ K) & ($10^{19}$ cm$^{-2}$)    & ($10^{16}$ cm$^{-2}$)     &($10^{15}$ cm$^{-2}$) & ($10^{13}$ cm$^{-2}$)}
\startdata
3C~273 & 139(92, 263) & 1.7(1.4, 2.0) &  3.5(2.3, 5.6)  & 2.0(1.3, 3.6) & 8.0(3.1, 14.7)  &6.0(3.8, 11.0)\\
Mrk~421 & 64(48, 104) & 1.4(1.3, 1.6) &  1.4(1.0, 2.0)  & 1.0(0.7, 1.5) & 1.5(0.6, ~2.6) &3.2(2.1, ~5.0)\\
GCSXE$^a$ & 216(104, 480) & 2.0(1.6, 2.3) &  2.2(1.4, 4.1) & 1.0(0.6, 2.6) & 7.2(2.0, 13.3)  &3.2(1.8, ~8.2)
\enddata
\tablecomments{In our absorption line model, the transition oscillation 
  coefficient and the damping factor are adopted from \citet{ver96}
  for \ovii\ and \oviii, and from  \citet{mor03} for \ovi. 
  90\% confidence intervals given in parenthesis.\\
  $^a$ The \xs\ 
  contribution along the sight line toward 3C~273 direction. See text for 
  the details.  
}
\end{deluxetable*}

\begin{figure} 
\plotone{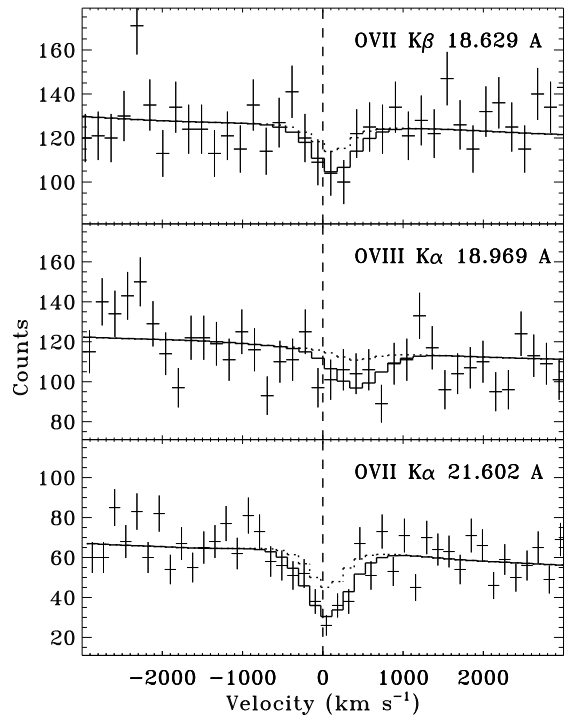}
\caption{Detected oxygen absorption lines in the {\sl Chandra}
	spectrum of 3C~273. The models have been convolved with the 
	instrumental responses. The dotted histograms
mark the amount of absorptions toward Mrk~421 direction.
The bin size is 10 m\AA.
   \label{fig:counts_MRK421} }
\end{figure}

Comparing with the Mrk~421 sight line, the 
3C~273 sight line is clearly affected by the hot gas associated with the \xs.
The model characterizations of the two sight lines
are substantially different, in terms of both thermal and kinematic properties 
as well as the total absorbing column densities (Table~\ref{tab:abs}).
We attribute this difference to the additional absorption produced by the 
\xs. We characterize this net \xs\ absorption by including a separate 
component in modeling the absorption along the 3C~273 sight line
(Fig.~\ref{fig:counts_MRK421}). 
This component is in addition to the reference contribution
assumed to be the same as that estimated from the Mrk~421 sight line
(Table~\ref{tab:abs}). Uncertainties in the reference contribution are neglected
because the spectral signal-to-noise ratio of Mrk~421
(e.g., Fig.~1 in Yao \& Wang 2007) is a factor of $\sim4.5$ 
higher than that of 3C~273.  The parameters of the \xs\ 
component are all free to be fitted. 

Similarly, we subtract the emission spectrum toward Mrk~421 from that 
toward 3C~273 to obtain the net emission spectrum of the \xs\ 
(Fig.~\ref{fig:ROSAT}). We have ignored the small difference in 
the foreground cool gas absorptions 
between the two sight lines (Table~\ref{tab:ROSAT}).
A correction for this difference will decrease
the emission intensity toward Mrk~421 by $\lsim5\%$, which is much less than
the statistical uncertainty (Fig.~\ref{fig:ROSAT}).

\begin{figure} 
\plotone{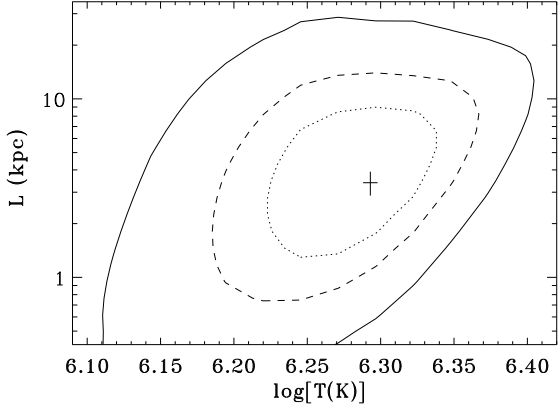}
\caption{\small The 68\%, 90\%, and 99\% confidence contours of the
effective scale length ($L$) vs. the temperature ($T$) of the hot 
gas associated with the X-ray emission enhancement toward 3C~273.
   \label{fig:L_vs_T} }
\end{figure}

We measure the properties of the hot gas associated with the 
X-ray enhancement toward 3C~273 by jointly fitting the net emission 
and absorption spectra. This joint fit, as good as that for the absorption 
data alone, matches the emission spectrum well 
(Figs.~\ref{fig:ROSAT}-\ref{fig:counts_MRK421}).
The fit gives $v_b$, $T$, and $N_{\rm H}$ and further allows us to infer 
$N_{\rm OVII}$, $N_{\rm OVIII}$, and $N_{\rm OVI}$ for the hot 
gas associated with the \xs\ (Table~\ref{tab:abs}). The same fit also
gives an estimate of the effective path-length through the gas as 
$L=3.4(1.0, 10.1)$ kpc. This inference uses the fact that the 
emission measure (EM) of the hot gas is proportional to its number density 
square ($EM\propto A_{\rm O} n_H^2L\eta$), whereas 
the line absorption depends on the 
ionic column density ($N_H\propto N_{\rm OVII}/A_{\rm O} \propto n_HL\eta$). 
For simplicity,  both the oxygen
abundance $A_{\rm O}$ (in solar units) and the volume filling factor 
$\eta$ are set to 1 in the quoted results (see \S~\ref{sec:dis} 
for further discussions). 
Figure~\ref{fig:L_vs_T} shows the $L$ versus $T$ confidence contours.
With these characterizations, we further infer the average density,
pressure, and EM of the gas as 
$n_H = 2.1(0.4, 6.6)\times10^{-3}~{\rm cm^{-3}}$,
$P/k = 4.22(0.8, 13.3)\times10^3~{\rm cm^{-3}~K}$,
and $EM=1.9(0.6, 6.3)\times10^{-2}~{\rm cm^{-6}~pc}$.

\section{Discussion}
\label{sec:dis}

We have obtained the dispersion velocity, temperature,
absorption column density, and the effective path-length of the diffuse gas 
associated with the X-ray emission enhancement toward the 3C~273 direction.
The 90\% lower limit of the path-length, $L\gsim1$ kpc, places the gas
well beyond any possible local origin within a distance of $D\lsim200$ kpc,
and therefore favors a remote origin of the gas.
The emission enhancement most likely reflects an outflow from the Galactic
center (GC) and/or bulge region. Large-scale winds driven by galactic nuclear 
starbursts and/or AGNs  have been observed 
in many galaxies (e.g., Martin 1999; Heckman \etal 2001). 
In our Galaxy, there is also evidence for the powerful energy injection 
from the GC on scales of several arcminutes to tens of degrees 
from radio to $\gamma$-ray 
(see Morris \& Serabyn 1996; Veilleux \etal 2005 for reviews).
The outflow is also expected from the energy and mass injection from
the Galactic bulge due to Type Ia supernovae and ejection of evolving 
low-mass stars. In any case, the outer boundary of the \xs\ may 
represent the interaction region of such an outflow with the 
Galactic disk and the accretion from the intergalactic medium
(Wang 2007). 

The large velocity dispersion ($v_b\sim216~{\rm km~s^{-1}}$) of the gas along
the 3C~273 sight line supports the GC origin of the \xs\ 
(Table~\ref{tab:abs}). Our estimate of the dispersion is based 
essentially on the relative saturation of the observed \ovii\ K$\alpha$
and K$\beta$ absorption lines and is consistent with the presence of
an \ovi\ high velocity absorption wing
(extending up to $\sim 240~{\rm km~s^{-1}}$), detected in the {\sl FUSE} 
spectrum of 3C~273 \citep{sem01}.
The dispersion is significantly higher than those estimated for the 
Mrk~421 sight line and for a similar off-\xs\ direction toward LMC~X--3 
(Galactic coordinates $l, b = 273\fdg57, -32\fdg08$; Fig.~\ref{fig:RASS};
Wang \etal 2005), indicating different natures of the intervening media.
At a temperature of $\sim2\times10^6$ K, the thermal broadening for the 
\ovii-bearing gas
is $\sim45~{\rm km~s^{-1}}$, which is negligible compared to 
$v_b$, indicating a large nonthermal broadening. 
This broadening cannot be due to the differential rotation of the Galactic 
disk, because of the high Galactic latitude direction and is most likely
due to the very turbulent nature of the gas, as is expected 
in the interaction region. 

It is also instructive to compare properties of the hot gas 
associated with the emission enhancement in the sight line of 3C~273 with 
those of the LHB. The temperature of the LHB is commonly accepted to be
$\sim10^6$ K, but the EM is still poorly constrained, 
ranging from 0.0079 to 0.02 ${\rm cm^{-6}~pc}$ (e.g., Smith \etal 2006; 
Galeazzi \etal 2007; Henley \etal 2007). Taking an average size of the LHB as 
100 pc (e.g., Sfeir \etal 1999), the EM implies an absorbing 
column density of $2.7-4.4\times10^{18}~{\rm cm^{-2}}$ across the LHB. 
The obtained column density of the hot gas associated with the enhancement, 
$\sim2.2\times10^{19}~{\rm cm^{-2}}$ (Table~\ref{tab:abs}), can therefore 
accommodate 
$\sim$ 5-7 bubble boundaries like those of the LHB in the line of 
sight, implying a size of the hot gas $\gsim500$ pc. The temperature of the 
hot gas is $\sim2$ times higher than that of the LHB (Table~\ref{tab:abs}).
Therefore, if the thermal pressure could be assumed to be roughly the same 
as that in the LHB, the size of the \xs\ would then be larger. This simple 
comparison also suggests that the \xs\ represents a remote phenomenon.

How does the conclusion depend on our assumed solar metal abundances of
the hot gas? We have used only the X-ray absorption lines produced by 
oxygen ions. For a plasma at $\sim2\times10^6$ K, \ovii\ and \oviii\ 
dominates ($\geq85\%$) the emission in the 
RASS 3/4-keV band. Therefore, a deviation of non-oxygen metal 
elemental abundances from the assumed solar values will not significantly 
change our $L$ estimate. For example, assuming the abundances 
equal to zero would give $L=2.8(0.9, 9.1)$ kpc, whereas adopting the
abundances to be supers-solar would lead to a slightly larger $L$.
The $L$ is nearly inversely proportional to the oxygen abundance.
To let the \xs\ be consistent with a local
origin (i.e., $L\lsim200$ pc; \S~\ref{sec:intro}) the oxygen 
abundance needs to be $\gsim5$ (90\% lower limit; \S~\ref{sec:results})
times solar, which is very unlikely.

With the characterizations of the hot gas obtained in \S~\ref{sec:results}, 
we can estimate the thermal energy
contained in wind material. We take a simple bipolar glass-timer 
geometry for the wind with a cone-like shape
near the GC and a cylindric shape at large vertical distances
($5\leq |z|\leq20$ kpc; e.g., Sofue 2000; Bland-Hawthorn \& Cohen 2003).
The radius of the cylinder is assumed to be $\sim7.2$ kpc, to be consistent
with our estimated path-length ($L \sim 3.4$ kpc) along the 3C~273 
sight line (taking the distance to the Galactic center of 7.6 kpc; 
Eisenhauer \etal 2005). To simplify our calculation, we assume 
that the gas density is uniformly distributed and the
gas temperature is exponentially decaying along the vertical direction from 
$5\times10^6$ K near the GC (e.g., Park \etal 1997) to $2\times10^6$ K at the 
latitude of 3C~273 (Table~\ref{tab:abs}). We find that total energy is 
$\sim10^{56}$ ergs, which is roughly in line with that expected in the
GC wind models \citep{sof00, bla03}.

While the emission enhancement toward 3C~273 likely arises in a region around
the GC, the bright NPS may still have a local origin 
(e.g., de Geus 1992). In this case, the NPS may be superposed on the 
large-scale \xs, which appears much more extended (to regions with $l \gtrsim
60^\circ$; Fig.~1). \citet{mil06} recently presented a high resolution 
{\sl Suzaku} emission spectrum of the bright NPS (Table~\ref{tab:ROSAT}).
From a multiple component fit, they found that while the relative 
abundances of Ne/O, Fe/O, and Mg/O are consistent with the solar values, 
the N/O is $\sim4$ times higher. They therefore attributed the NPS to
the Sco-Cen environment, being enriched by material 
undergoing the CNO cycle in massive stars. In addition,
the temperature they obtained,  $\sim3.4\times10^6$ K,
is substantially higher than that of the \xs, as we have obtained here
(Table~\ref{tab:abs}). A similarly high temperature was also obtained 
from the {\sl XMM-Newton} observations \citep{wil03}. The apparent high 
temperature of the NPS may be due to the high-velocity shock-heating, 
commonly seen in young supernova remnants. 

We have assumed that the hot gas associated with the \xs\ along the
3C~273 sight line is isothermal, which is adequate in the present study 
due to the very limited spectral resolution of the RASS data used here. 
Emission data with a low instrument background and relatively high
spectral resolution (e.g., from a deep {\sl Suzaku} X-ray CCD observation)
will be particularly useful for tightening the constraints on the properties of 
the hot gas associated with the \xs.

\acknowledgements 
We thank Eric Miller, Claude Canizares, Paola Testa, and 
Herman Marshall for useful discussions. We are also grateful to 
the anonymous referee 
for insightful comments and suggestions, which helped to 
improve the presentation of the paper.
This work is supported by NASA through the 
Smithsonian Astrophysical Observatory contract SV3-73016 to 
MIT for support of the {\sl Chandra} X-Ray Center under contract NAS 08-03060. 
Support from {\sl Chandra} archival research grant AR7-8016 are also 
acknowledged. QDW appreciates additional support from NASA grant NNX06AB99G.

\clearpage

\clearpage

\end{document}